\documentclass[reprint, amsmath,amssymb, aps, pra, longbibliography]{revtex4-1}

\usepackage{soul}
\usepackage{amsmath}   
\usepackage{amssymb}
\usepackage{graphicx}
\usepackage{dcolumn}
\usepackage{bm}
\usepackage{amsfonts}
\usepackage{subfigure}
\usepackage{array}
\usepackage{float}
\usepackage{color}
\usepackage[colorlinks=true,linkcolor=blue]{hyperref}%
\hypersetup{allcolors=blue}
\usepackage[normalem]{ulem}
\usepackage{xcolor}

\newcommand{\ket}[1]{\ensuremath{\left\vert #1 \right\rangle}}

\begin{document}

\title{Spectroscopy of the $^{85}$Rb 4$D_{3/2}$ state for hyperfine-structure determination}
\date{\today }

\author{A. Duspayev}
    \email{alisherd@umich.edu}
\affiliation{Department of Physics, University of Michigan, Ann Arbor, MI 48109, USA}
\author{G. Raithel}
\affiliation{Department of Physics, University of Michigan, Ann Arbor, MI 48109, USA}

\begin{abstract}
We report a measurement of the hyperfine-structure constants of the $^{85}$Rb 4$D_{3/2}$ state using a two-photon 5$S_{1/2}\rightarrow$4$D_{3/2}$ transition. The hyperfine transitions are probed by measuring the transmission of the low-power 795-nm lower-stage laser beam through a cold-atom sample as a function of 795-nm laser frequency, with the frequency of the upper-stage 1476-nm laser fixed. All 4 hyperfine components are well-resolved in the recorded transmission spectra. AC shifts are carefully considered. The field-free hyperfine line positions are obtained by extrapolating measured line positions to zero laser power. 
The magnetic-dipole and electric-quadrupole constants, $A$ and $B$, are determined from the hyperfine intervals to be 7.419(35)~MHz and 4.19(19)~MHz, respectively.
The results are evaluated in context with previous works. Possible uses of the Rb 4$D_J$ states in Rydberg-atom-physics, precision-metrology and quantum-technology applications are discussed.   
\end{abstract}

\maketitle

\section{Introduction}
\label{sec:intro}
Transitions between internal energy states of atoms and molecules and their properties form a basis for many modern applications in fundamental science~\cite{safronovarmp} and emerging quantum technologies~\cite{macfarlane, Adams_2020, weiss2017}. Prominent examples include sub-Hz-linewidth transitions in atomic clocks~\cite{bloom2014}, shifts (or the absence thereof) of transitions used in quantum gates or laser traps~\cite{SafronovaRbMagic, Lundblad2010, zhangpra2011, sahoo2013}, and transitions with wavelengths in optical communications bands~\cite{chanliere2006, Cao_2019, Menon_2020}. At the same time, precision measurements are an important test bed for atomic-structure calculations~\cite{SAFRONOVA2008191, allegrini} and searches for novel physics~\cite{safronovarmp}.

Low $D$ states of rubidium have been a subject of interest due to several reasons. Two-photon transitions into these states from the ground 5$S_{1/2}$ state are relatively strong and can be driven using readily available lasers at visible and NIR wavelengths. A third laser, also usually at a wavelength accessible with a diode laser, can be added to excite the $D$-state atoms further-up into Rydberg states~\cite{thoumany2009, Fahey11, johnson2012, lim2022} for studies on electromagnetically induced transparency~\cite{carr2012, Moore2019a}, Rydberg molecules~\cite{shafferreview, feyreview, duspayev2021, deiss2021, zuber2022}, excitation of states with orbital quantum number $\ell > 3$~\cite{Younge2010, moore2020, cardman2021njp}, and preparation of circular Rydberg atoms~\cite{cardman2020, wu2023}. The Rb $5S_{1/2} \rightarrow$ $4D_J$ and $5S_{1/2} \rightarrow$ $5D_J$ two-photon transitions are also quite narrow, making them attractive for realizing novel frequency standards~\cite{hilico1998epjapmetrological, quinn2003metpractical, moon2004, terra2016apbultra, Roy17}. In particular, there is an ongoing effort to utilize the Rb 5$D_J$ states in robust, portable atomic clocks~\cite{martin2019}. Measurements on Rb 5$D_J$ states in 1064~nm dipole traps~\cite{duncan2001prameasurement, cardman2021} have revealed large photo-ionization (PI) cross sections. PI of atoms laser-cooled and -trapped in a NIR-laser focus may be a convenient method to prepare a cold-ion source~\cite{ionsourcepaper}. In Rydberg-physics applications that involve 
Rydberg-atom excitation in 1064-nm (or similar) laser traps via a Rb $5D_J$-state, PI presents an unwanted complication, even at moderate 1064-nm laser power~\cite{cardman2021, atoms10040117}.
The Rb 4$D_J$ state, however, has an ionization wavelength of $698~$nm, allowing the use of NIR laser traps in the above applications without limitations due to PI. Moreover, the two-photon transitions into Rb $4D_J$-levels via the D-lines involve lasers at telecom wavelengths, offering opportunities for applications in quantum communication~\cite{chanliere2006, Cao_2019, Menon_2020}. Thus, precision measurements of the properties of Rb $4D_J$-levels, including their hyperfine structure (HFS)~\cite{rmp1977, allegrini}, are of interest in the aforementioned research directions.

Several recent studies report investigations of the 4$D_{5/2}$ HFS~\cite{Lee_07, wang2014, lee2015, Roy17}. For the 4$D_{3/2}$ state, only one frequency interval between its hyperfine states has been measured~\cite{liao1974, moon2009}. The HFS measurements in these studies were performed in room-temperature Rb vapors. Other recent work on the E2 transition $5S_{1/2} \rightarrow$ $4D_{3/2}$ has employed cold atoms~\cite{Ray_2020}. Here we report measurements of all three frequency intervals of the 4$D_{3/2}$ HFS using laser-cooled and -trapped Rb atoms. This allows us to extract the HFS constants, $A$ and $B$, without having to resort to previously-measured HFS constants of other atomic levels~\cite{liao1974, moon2009}. Our paper is organized as follows: the experimental setup, data acquisition and data analysis methods are described in Sec.~\ref{sec:methods}. Our uncertainty analysis includes a careful consideration of all AC Stark shifts. Results are presented, discussed and compared to previous works in Sec.~\ref{sec:res}. We also discuss future possible applications of the Rb 4$D_J$ states in metrology and quantum technologies. The paper is concluded in Sec.~\ref{sec:concl}.

\section{Methods}
\label{sec:methods}

\subsection{Theory of HFS}
\label{subsec:theory}
The HFS is a result of the coupling between the total electronic and the nuclear angular momenta, $\textbf{J}$ and $\textbf{I}$, respectively~\cite{footbook}. The energy splittings between neighboring hyperfine states with hyperfine quantum numbers $F$ and $F-1$ can be expressed as

\begin{equation}
\nu_{F}- \nu_{F-1} = A F + B \frac{\frac{3}{2} F (F^2 - I(I+1) - J(J+1) + \frac{1}{2})}{I (2I-1) J (2J - 1)},
\label{eq:hfsplit}
\end{equation}

\noindent where $J$ and $I$ are the electronic and nuclear angular-momentum quantum numbers, respectively ($I = 5/2$ for $^{85}$Rb and $J = 3/2$ for the 4$D_{3/2}$ state), and $A$ and $B$ are the magnetic-dipole and electric-quadrupole HFS constants. In Eq.~\ref{eq:hfsplit} the magnetic-octupole HFS interaction, commonly associated with a HFS constant labeled $C$~\cite{SteckRb85}, is neglected because it is too small to be measured.   

From now on we will concentrate on the $^{85}$Rb 4$D_{3/2}$ state. For the hyperfine quantum numbers of the atomic states used in our excitation scheme shown in  Fig.~\ref{fig1}~(b), we use the labels $F$ for the ground state 5$S_{1/2}$, $F'$ for the intermediate state 5$P_{1/2}$, and $F''$ for 4$D_{3/2}$. 
We determine the HFS constants $A$ and $B$ of 4$D_{3/2}$ from the gaps $\nu_{F''}-\nu_{F''-1}$ using Eq.~\ref{eq:hfsplit}, as described in Sec.~\ref{subsec:data}.

\subsection{Experimental setup}
\label{subsec:expsetup}

\begin{figure}[t!]
 \centering
  \includegraphics[width=0.48\textwidth]{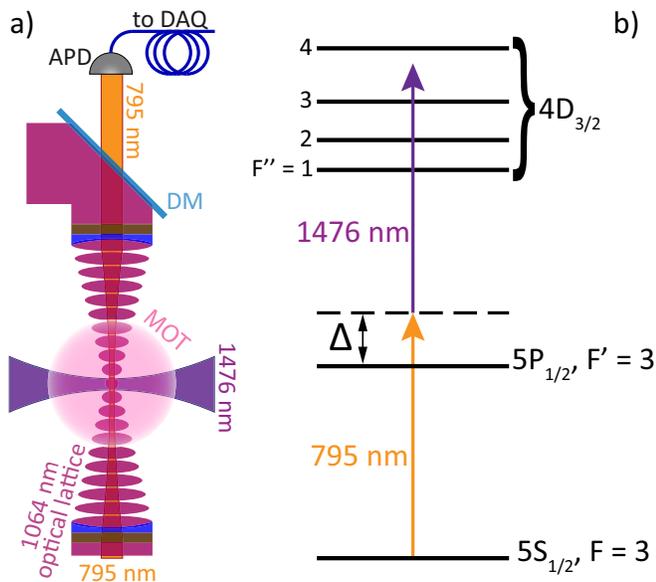}
  \caption{(Color online) (a) Most relevant components of 
  the experimental setup. ``APD" avalanche photo-detector; ``PD" photo-diode; ``DM" dichroic mirror;
  ``MOT" magneto-optical trap. See text for details.
  (b) Utilized laser excitation scheme (not to scale). } 
  \label{fig1}
\end{figure}

The main aspects of the utilized experimental setup are depicted in Fig.~\ref{fig1}~(a). $^{85}$Rb atoms are laser-cooled and trapped in a 3D magneto-optical trap (MOT). As this happens, they are simultaneously loaded into an intracavity 1064-nm optical lattice (OL) of depth $U_{latt} \sim h \times 4$~MHz for 5$S_{1/2}$-atoms, allowing us to achieve an atom density of $\lesssim 10^{11}$~cm$^{-3}$. The main features of the setup's design and of the atom preparation method have been described in detail elsewhere~\cite{Chen2014praatomtrapping}. In recent experiments performed with this setup~\cite{cardman2021,atoms10040117}, the atomic response to 1064-nm light has been the central subject of study. In the present measurements the role of the 1064-nm OL is limited to providing an atom column density that is high enough for the signal-to-noise ratio of the atomic absorption lines of interest to reach a level at which the measurements become practical.

Each experimental cycle (repetition rate of 100~Hz) begins with close to 10~ms of collecting atoms in the MOT and in the OL.
After loading atoms into a dense column (diameter $\sim 20~\mu$m, length $\gtrsim 500~\mu$m) along the OL axis,
the lasers for MOT and OL are switched off, while the MOT repumper beam is left on. The two-photon excitation, 5$S_{1/2}\rightarrow$4$D_{3/2}$, is performed using 795~nm and 1476~nm lasers [see Fig.~\ref{fig1}~(b)]. The former is frequency-controlled using an optical phase-locked loop (OPLL), 
which locks the measurement laser at a well-controlled frequency offset relative to a different 
laser that is locked to the 5$S_{1/2}, F=3 \rightarrow$5$P_{1/2}, F'=3$ transition using saturated spectroscopy. The 795-nm measurement beam is co-aligned with the OL axis, as it propagates through the chamber, ensuring good spatial overlap with the high-density column of atoms prepared by the OL. The 795-nm measurement beam has a power of $\sim 50$~nW before the OL cavity and a waist of $\gtrsim$~20~$\mu$m around its center (where the atoms are collected in the OL) and is pulsed on for 2~$\mu$s while MOT and OL lasers are switched off.
The 1476-nm laser falls onto the atoms from the side, as shown in Fig.~\ref{fig1}~(a). Efficient spatial overlap between the 1476-nm light and the high-density column of cold atoms is achieved using a cylindrical beam-shaping lens before the vacuum system [not shown in Fig.~\ref{fig1}~(a)]. The 
1476-nm laser is locked at a fixed frequency to a temperature-stabilized Fabry-P\'erot etalon, is kept always on, and is linearly polarized along the cavity axis. Since 
the laser beams entering the OL cavity are linearly polarized as well~\cite{atoms10040117}, the two-photon excitation has a  lin. $\perp$ lin. polarization configuration.

As the upper-transition (1476-nm) laser is fixed in
frequency, we scan the two-photon 5$S_{1/2} \rightarrow$ $4D_{3/2}$ transition by scanning the 795-nm laser via the frequency control afforded by the OPLL lock~\cite{cardman2021}. Hence, the intermediate-state detuning, $\Delta$, defined to be 0~MHz at the field-free $\ket{5S_{1/2}, F = 3} \rightarrow \ket{5P_{1/2}, F' = 3}$ transition [see Fig.~\ref{fig1}~(b)], and the two-photon detuning of the  5$S_{1/2} \rightarrow$ $4D_{3/2}$ transition are scanned synchronously. The transmission of the 795-nm laser through the cylindrical atomic sample, $T_{795}$, is then measured using an avalanche photo-detector (APD) located after the output port of the OL cavity [see Fig.~\ref{fig1}~(a)]. We scan $\Delta$ over a fairly far-detuned range $\Delta \gtrsim 150$~MHz to avoid line broadening and extra spectral lines due to population of the intermediate 5$P_{1/2}$ level, and to minimize the relative variation in $\Delta$ across the range within which the 4$D_{3/2}$ HFS resonances occur (which is about 70~MHz wide). Absorption spectra are recorded for a selection of powers of the upper-transition laser, $P_{1476}$, which is varied between $\sim$0.5~mW and 5~mW, as a function of $\Delta$. To improve statistics, the spectra are averaged over 5 identical scans at high $P_{1476}$ to 30 identical scans at low $P_{1476}$.

\subsection{Transmission spectra, HFS line shifts, and systematic effects}
\label{subsec:hfsresult}

\begin{figure}[t!]
 \centering
  \includegraphics[width=0.48\textwidth]{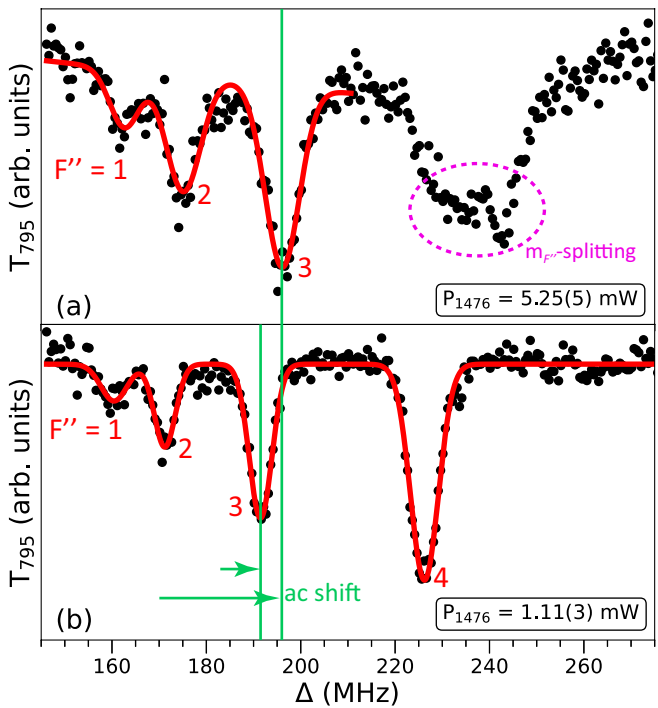}
  \caption{(Color online) Examples of measured transmission spectra for two power settings of the 1476~nm laser, $P_{1476}$. Dots show experimental data averaged over several scans. Solid lines show multi-peak Gaussian fits. The AC Stark shift 
  caused by the 1476~nm laser light is visualized on the $F''=3$-peak.
  The $F''=4$ line at $P_{1476}=5.2$~mW has several marginally resolved sub-components that are split due to the dependence of the AC shift on the magnetic quantum number $m_{F''}$.} 
  \label{fig2}
\end{figure}

Two examples of sample-averaged transmission spectra, $T_{795}$ vs $\Delta$, at two different indicated settings of $P_{1476}$ are shown in Figs.~\ref{fig2}~(a) and (b). We can clearly resolve all 4$D_{3/2}$ $F''=1$ to 4 resonances. We determine the line centers with sub-MHz resolution using multi-peak Gaussian fits, with the exception of the $F''=4$-line at the highest
powers $P_{1476}$. Generally, the $F'' = 4$ state is the strongest in all of the acquired averaged spectra. The $F''=4$ and $F''=3$-peaks remain visible even at the lowest $P_{1476}$ used, while the $F'' = 1$ and $F''=2$-peaks become almost indiscernible from the background noise on $T_{795}$. The lowest achieved linewidth is $<$~5~MHz at the lowest $P_{1476}$, which is slightly smaller than in a previous study~\cite{moon2009}, but still larger by a factor $>$~2 than the estimated natural linewidth of the 4$D_{3/2}$ state (which is 2.02~MHz, corresponding to a lifetime of $78.7$~ns~\cite{Heavens_61}). The 
extra width results from the $m_{F''}$-dependence (and likely some inhomogeneity) of the AC shifts, interaction time broadening (200~kHz), laser linewidth (a few 100~kHz), and symmetric Zeeman broadening due to the MOT magnetic field ($\lesssim 1$~MHz). 
Saturation broadening does not play a role.
All broadening effects except the AC shift contribute to a symmetric widening of the lines that does not shift the line centers and merely results in added statistical uncertainty. The AC shifts from the 1476-nm laser field are evident in Fig.~\ref{fig2}, where the AC shifts increase by up to about 18~MHz when $P_{1476}$ is increased from 1.1~mW to 5.2~mW. The AC shifts depend on $F''$ and $m_{F''}$. In all cases shown except one, the
$m_{F''}$-splitting is not resolved; in these cases the lines appear AC-shift-broadened. The $m_{F''}$-splitting of the AC shift is marginally resolved at 5.2~mW and $F''=4$.

\begin{figure}[htb]
 \centering
  \includegraphics[width=0.48\textwidth]{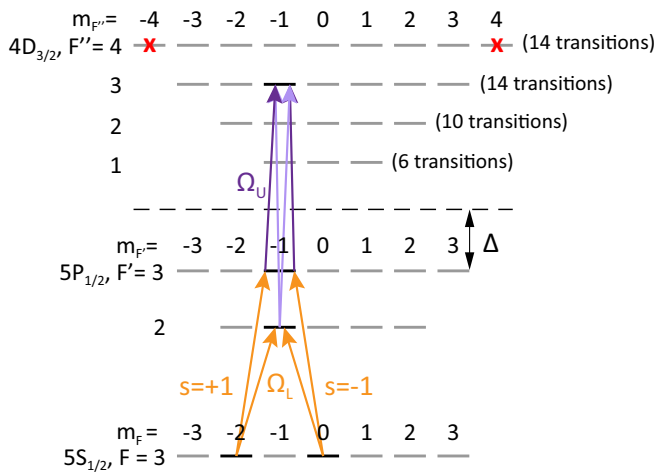}
  \caption{(Color online) Expanded energy-level diagram (including magnetic substructure) of the two-photon transitions shown in Fig.~\ref{fig1}~(b). For clarity, excitation channels are shown only for the target state $|F''=3, m_{F''}=-1 \rangle$ (a random pick out of many). The levels relevant to this case are printed in a darker shade. Transition amplitudes must be coherently summed over $F'$ to result in two-photon Rabi frequencies $\Omega(F'',m_{F''}, s)$, as seen in Eq.~(\ref{eq:2phrabi}).
  The crossed-out levels are not accessible with the laser polarizations used.
  } 
  \label{fig3}
\end{figure}

In the utilized lin. $\perp$ lin. polarization scheme, and using the electric-field direction of the 1476-nm beam as quantization axis, the magnetic substructure of the two-photon transitions follows the scheme
shown in Fig.~\ref{fig3}. Due to the magnetic isotropy of the atom sample, the absence of optical pumping, and the absence of Zeeman coherences in the ground state, the lines can be viewed as an incoherent superposition of 14, 14, 10 and 6 lines for $F''=4$, 3, 2 and 1, respectively, as seen in Fig.~\ref{fig3}, with equal populations in all ground $m_F$-sublevels. 
We denote the lower- and upper-transition Rabi frequencies through the
intermediate $F'=2$ and $F''=3$-levels as $\Omega_{L} (F',m_{F''}, s)$ and $\Omega_{U} (F',F'',m_{F''})$, where the identifier $s=+1$, $-1$ is used as an index for $\sigma^+$- and $\sigma^-$-excitation on the lower transition (the laser polarization direction of which is along the $x$-axis). The hyperfine shifts of the
$5P_{1/2}$ and $4D_{3/2}$-levels, denoted
$\Delta_{HFS}(F')$ and $\Delta_{HFS}(F'')$, are referenced to the respective uppermost hyperfine levels (so that all values are zero or negative). For the chosen polarizations, $m_{F'}=m_{F''}$ and $m_F=m_{F'}-s$. The two-photon Rabi frequencies then are 
\begin{align}
\label{eq:2phrabi}
\begin{split}
\Omega\ &(F'',m_{F''}, s) = \\
&\sum_{F'} \big[ \Omega_{L} (F',m_{F''}, s) \, \Omega_{U} (F',F'',m_{F''}) \Big] / \\ 
&\big[2 (\Delta + \Delta_{HFS}(F'')  + \delta\nu_{AC}(F'',m_{F''}) \\ 
&-  \Delta_{HFS}(F')  -  \delta\nu_{AC}( F',m_{F''}) )\big] \quad,
\end{split}
\end{align}
where the AC shifts are denoted $\delta\nu_{AC}(F'',m_{F''})$ and $\delta\nu_{AC}(F',m_{F''})$. The sum has a maximum of two terms. 
Our computation of the two-photon Rabi frequencies $\Omega(F'',m_{F''}, s)$ 
yields values in the range of $\lesssim 2 \pi \times 0.6$~MHz, peaking at $2 \pi \times 0.6$~MHz for transitions into $\vert F''=4, m_{F''} = 0 \rangle$ for the highest beam powers, and for the beam sizes and detunings $\Delta$ used. Hence, the two-photon transitions are not significantly saturated in the power range in Fig.~\ref{fig2}. Therefore, the two-photon line strengths measured in terms of single-atom photon scattering rates are
\begin{equation}
S(F'',m_{F''}) \approx \sum_s \frac{ [ \Omega(F'',m_{F''}, s) ]^2}{\Gamma_{4D}} \quad,
\end{equation}
with the $4D_{3/2}$ decay rate $\Gamma_{4D} = 2 \pi \times 2.02$~MHz. For our beam parameters and $\Delta_0= 225$~MHz, the computed line-strength ratios summed over $m_{F''}$ are
about 1 : 0.53 : 0.33 : 0.12 for $F''$=4, 3, 2 and 1, in good qualitative agreement with respective ratios from Fig~\ref{fig2}~(b), measured at low 1476-nm power, 1 : 0.49 : 0.29 : 0.08. There, we define $\Delta_0$ as the detuning $\Delta$ at which the $F''=4$-resonance occurs in the limit of zero laser power. In the experiment, the value of $\Delta_0$ is set by the 1476-nm laser, which is held at a fixed frequency during the 795-nm-laser scan. 

The light shifts can be separated into far-off-resonant contributions and
near-resonant shifts. We have computed the polarizabilities of the three relevant levels with the near-resonant terms peeled out, and have estimated that the far-off-resonant shifts of the $5S_{1/2}$, $5P_{1/2}$ and $4D_{3/2}$-levels are below 100~Hz, 150~Hz and 1~kHz, respectively, and can therefore be safely ignored. The positive shift of the $5S_{1/2}$-level due to the near-resonant but low-intensity 795-nm beam is estimated at 200~kHz, which is in range of our measurement precision  $(\sim 100~$kHz). 
However, because we extract the hyperfine constants $A$ and $B$ from the differences between 4$D_{3/2}, F''$ line positions, the common-mode shift due to the
$5S_{1/2}$ AC Stark effect from the 795-nm laser field 
drops out. Hence, in the present work we ignore the small AC shift of the $5S_{1/2}$-level in the 795-nm laser field. The shift of the $5P_{1/2}$-level due to the 795-nm beam can be safely ignored because the intermediate states are off-resonant by $\gtrsim$200~MHz [see Fig.~\ref{fig1}~(b)] in our two-photon excitation scheme.

The light shifts of the $4D_{3/2}$-levels due to the 1476-nm beam,
\begin{equation}
\delta\nu_{AC}(F'',m_{F''}) = \sum_{F'} \frac{|\Omega_{U} (F',F'',m_{F''})|^2}{4 (\Delta + \Delta_{HFS}(F'') - \Delta_{HFS}(F'))} \quad,
\end{equation}
are in the range of tens of MHz and cause line shifts proportional to 1476-nm power. 
In our experiment, the AC shift of the $4D_{3/2}$ sublevels peaks at about 20~MHz for $\vert F''=4, m_{F''} = 0 \rangle$, which also has the largest two-photon Rabi frequency and therefore contributes the most to the $F''=4$-line. 
The light shifts of the $5P_{1/2}$-levels due to the 1476-nm beam follow from an analogous equation, peak at about -27~MHz for $\vert F'=3, m_{F''} \pm3 \rangle$, and merely cause a small change in line strengths, as seen in Eq.~\ref{eq:2phrabi}.
We have confirmed that the light-shift amounts experimentally observed in Fig.~\ref{fig2} are in accordance with our estimated intensities of the 1476-nm beam. 
The calculation further shows that the light shifts generally increase with the value of $F''$, as also seen in the experimental data in Fig.~\ref{fig4} below. The $m_{F''}$-dependence of $\delta\nu_{AC}(F'',m_{F''})$ causes most of the substructure seen in the $F''=4$-line in Fig~\ref{fig2}~(a), which has the largest AC shifts. The maximum AC shift of $F''=4$, seen in Fig.~\ref{fig2}~(b), allows us to perform a rough calibration of the 1476-nm electric field, which yields 3500~V/m at the highest powers used. 

The MOT magnetic fields are $\lesssim 0.5$~Gauss and cannot be turned 
off because the MOT field is generated with permanent magnets. Because the atomic sample is not magnetized, and because the MOT field near the MOT trapping region tends to be randomly oriented, when averaged over all atoms, the Zeeman shifts are symmetric and merely cause a line broadening of up to about 1~MHz, with no net line shift at our level of precision $(\sim 100~$kHz). 
It is noted that in previous work~\cite{cardman2021, atoms10040117} on this setup
there were also no observable effects from Zeeman shifts.
At the highest 1476-nm power used, the strong and maximally AC-shifted $|F=3, m_F= \pm 1 \rangle$ $\rightarrow$ $|F''=4, m_F= 0 \rangle$ line, which is less magnetic-field-broadened than most other magnetic sub-transitions,
likely causes the relatively sharp feature at the very right of the spectrum in Fig.~\ref{fig2}~(a). 

For completeness, we add that DC Stark effects are irrelevant 
for low-lying atomic states in our system, in which DC electric fields have been eliminated via Stark spectroscopy of Rydberg levels (which serve as highly sensitive electric-field sensors~\cite{ionsourcepaper}). Similarly, density shifts of the relevant states are not significant at our level of precision. Accordingly, in the experiment we have not observed any line shifts that depend on MOT atom density or on the effectiveness of the OL used to accumulate atoms in a dense column.  

Computations and experimental data both show that for $\Delta_0=225$~MHz the two-photon absorption in our cold-atom sample is only a few percent. The off-resonant one-photon absorption on the D1-line adds additional absorption of a few tenths of a percent. The 795-nm laser power measured as a function of 795-nm laser frequency changes linearly by several percent due to the current feed-forward used in the 795-nm laser scan. To avoid line pulling, we subtract the linear background from the spectra before fitting. 

Summarizing our analysis of line structures and shifts, we conclude that with the
exception of the AC Stark shift our measurements are free of significant systematic effects.
Importantly, as the 795-nm laser is scanned via an OPLL with sub-kHz precision and accuracy across the entire scan range, the frequency calibrations in Figs.~\ref{fig2} and~\ref{fig4} are perfectly linear and uncertainty-free at our level of precision. Hence, the only systematic effect that must still be addressed in the data analysis is the AC Stark shift from the 1476-nm laser.

\subsection{Data analysis}
\label{subsec:data} 

We have obtained spectra as in Fig.~\ref{fig2} for a selection of seven 1476-nm laser powers, $P_{1476}$. All spectra show resolved $F''$ lines, the centers of which are determined by multi-peak Gaussian fits. At the highest $P_{1476}$, we do not fit the $F''=4$-line because of its obvious magnetic substructure. Further, at the lowest powers we do not fit the lowest-$F''$ lines because they are too noisy.

For the determination of the 4$D_{3/2}$ HFS constants, the $F''$ line positions for $P_{1476} \rightarrow 0$ are required. To that end, we plot the line positions from the multi-Gaussian fits to the spectra as a function of $P_{1476}$, separately for each $F''$.
Noting that the AC shifts are linear in  $P_{1476}$, the zero-power y-intercepts of linear fits to the data reveal the desired 
$F''$ line positions. In this way, the state-dependent AC Stark shifts are corrected for. Further, the fit uncertainty of the
y-intercepts serves as the (only relevant) systematic uncertainty
in our experiment. We then use the zero-power intercepts to obtain the frequency intervals between adjacent $F''$-states, $\nu_{F''} - \nu_{F''-1}$. Subsequently we extract the HFS constants $A$ and $B$ using two methods. 

\textit{Method 1.} The most straightforward way to extract $A$ and $B$ is to directly solve the system of linear equations derived from Eq.~\ref{eq:hfsplit} for a pair of $\nu_{F''}-\nu_{F''-1}$-values, and to propagate the measurement uncertainties to determine the uncertainties of $A$ and $B$~\cite{taylor}. As there are two unknowns and three frequency intervals, there is a redundancy. However, as discussed in Sec.~\ref{sec:res}, the rather high uncertainty on the lowest interval, $\nu_{2} - \nu_1$, makes that interval measurement near-irrelevant. 
Therefore, in Method 1 we only utilize the intervals
$\nu_4 - \nu_3$ and $\nu_3 - \nu_2$, allowing us to solve for both 
$A$ and $B$. 

\textit{Method 2.} An alternative approach is to fit Eq.~\ref{eq:hfsplit} to the experimental data for all three $\nu_{F''} - \nu_{F''-1}$ intervals, with $A$ and $B$ as fitting parameters, regardless of the large uncertainty of the lowest interval. 
This results in uncertainties on $A$ and $B$ that are larger than in Method 1 by a factor $\gtrsim 3$. Therefore, we use this approach only for a consistency check. The results from Method 1 are reported as the final result of our measurements.

\section{Results and Discussion}
\label{sec:res}

\subsection{HFS line positions}
\label{subsec:lines}

\begin{figure}[htb]
 \centering
  \includegraphics[width=0.48\textwidth]{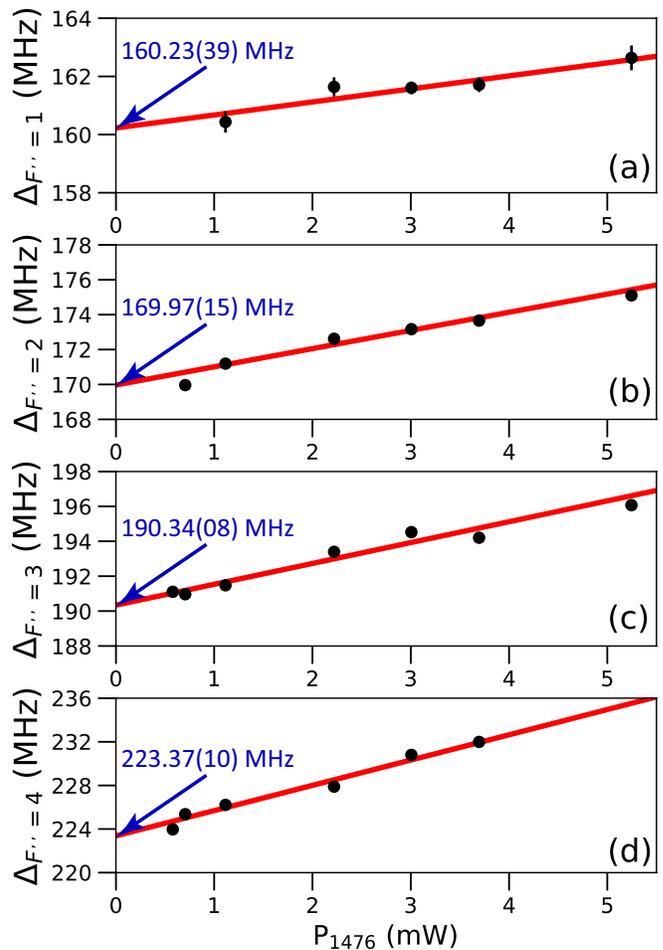}
  \caption{(Color online) Line centers of the 4$D_{3/2}$ HFS components, $\Delta_{F''}$, as a function of 1476-nm laser power, $P_{1476}$, 
  for $F''=1$ (top panel) to $F''=4$ (bottom panel). The dots show results from multi-peak Gaussian fits to experimental spectra (for specific examples see Fig.~\ref{fig2}). Vertical error bars are the fitting uncertainties, and horizontal error bars are from the $P_{1476}$ measurement. Most error bars are too small to be seen.
  The solid lines are linear fits with y-intercepts and uncertainties as indicated.} 
  \label{fig4}
\end{figure}

\begin{center}
\begin{table}[htb]
\caption{\label{tab:table1} Frequency intervals between neighboring $^{85}$Rb 4$D_{3/2}$ $F''$ states obtained from the zero-field intercept
method presented in Fig.~\ref{fig4}.}
\begin{tabular}{|c| c|}
    \hline
    $F''$ & [$\nu_{F''}-\nu_{F''-1}$]~(MHz)\\
    \hline
    4 & 33.0(1)\\
    3 & 20.4(2)\\
    2 & 9.7(4)\\
    \hline
\end{tabular}
\end{table}
\end{center}

In Fig.~\ref{fig4} we plot the fitted line centers, denoted as $\Delta_{F'' = i}, i = 1, 2, 3, 4$, versus $P_{1476}$ for $F''=1$ (top) to 4 (bottom). The zero-field energy levels of the HFS states are the extracted by fitting the data by linear functions. The fit results are shown in Fig.~\ref{fig4} as solid lines. The extracted y-intercepts yield the desired zero-field level positions. As expected, we find that in the cases in which we are able to measure the line centers down to the lowest 1476-nm laser powers the zero-field intercepts have the lowest uncertainties [$F'' = 3$ and 4 in Figs.~\ref{fig4}~(c) and (d), respectively]. The determined frequency intervals $\nu_{F''} - \nu_{F''-1}$ are listed in Table~\ref{tab:table1}. As described in Sec.~\ref{subsec:data}, these are used to determine the $A$ and $B$ HFS constants using the two methods
in Sec~\ref{subsec:data}. The results of these analyses are summarized in Table~\ref{tab:table2}.

\subsection{HFS constants and comparison with previous results}
\label{subsec:compare}

\begin{table}[htb]
\caption{\label{tab:table2} $A$ and $B$ HFS constants determined in this study and comparison with previous reports. The corresponding uncertainties are in brackets. M1 and M2 stand for Method 1 and 2 explained in Sec.~\ref{subsec:data}, respectively. All values are in MHz.}
\begin{tabular}{|c |c |c |c |c |}
    \hline
    Constant & \cite{liao1974} & \cite{moon2009} & This work, M1 & This work, M2\\
    \hline
    A & 7.3(5) & 7.329(35) & 7.419(35) & 7.37(12)\\
    B & - & 4.52(23) & 4.19(19) & 4.48(63)\\
    \hline
\end{tabular}
\end{table}

The extracted $A$ and $B$ constants from Methods 1 and 2 (M1 and M2, respectively, in Table~\ref{tab:table2}) agree with each other within their uncertainties. 
The uncertainties in Method 2 are larger by factors $>$~3. This is due to the large
uncertainty of the lowest HFS gap, which in turn is due to the low signal strength of the $F''=1$ line, negating a determination of the $F''=1$ line centers at the lowest powers $P_{1476}$ and causing a large uncertainty of the y-intercept for $F''=1$. This results in a large 
uncertainty of the gap $\nu_2-\nu_1$.    
Whereas, in Method 1 we only use the gaps with low uncertainties, resulting in more precise values for $A$ and $B$.
Nevertheless, we leave the results from Method 2 as a consistency check and report the results from Method 1 as final:

\begin{eqnarray}
A & = & 7.419(35)~{\rm{MHz}}\nonumber \\
B & = & 4.19(19) ~{\rm{MHz}}. \nonumber
\end{eqnarray}

We note that our results for $A$ and $B$ are extracted as independent variables or fit parameters from our data. In contrast, in previous measurements only one HFS frequency interval was measured, and previously-known ratios of the HFS constants of the two Rb isotopes for the different energy states (5$S_{1/2}$ and 5$P_{3/2}$) had to be brought in for the determination of both $A$ and $B$ for 4$D_{3/2}$~\cite{moon2004, moon2009}. 
Specifically, the assumption that the ratio of the HFS constants of different states for different isotopes ($^{85}$Rb and $^{87}$Rb) should be same has been used. As in our work we are able to observe 3 out of 4 hyperfine states with high resolution, our analysis is independent from any previous studies and  model-specific assumptions.

We further note that our result for $A$ lies outside of the uncertainty overlap compared to the most recent measurement~\cite{moon2009}, with both values having almost same uncertainty, whereas our result for $B$ agrees with~\cite{moon2009}, with our value having slightly smaller uncertainty. The difference between the $A$-values is most probably a consequence of the different approaches to the HFS determination.

It is noteworthy that our result for $\nu_{4} - \nu_3$ 
in Table~\ref{tab:table1} agrees well with the value from~\cite{moon2009}. Empirically we find that an artificial decrease of the lower frequency intervals reported in our work, that were not observed in~\cite{moon2009}, can lead to a better agreement between the results for the HFS constants. The borderline-significant deviation seen in $A$ merits additional experimental and theoretical investigations that may be useful for questions in fundamental physics, such as hyperfine anomaly in different isotopes~\cite{wang2014} and parity nonconservation~\cite{roberts2014}.

\begin{figure}[t!]
 \centering
  \includegraphics[width=0.48\textwidth]{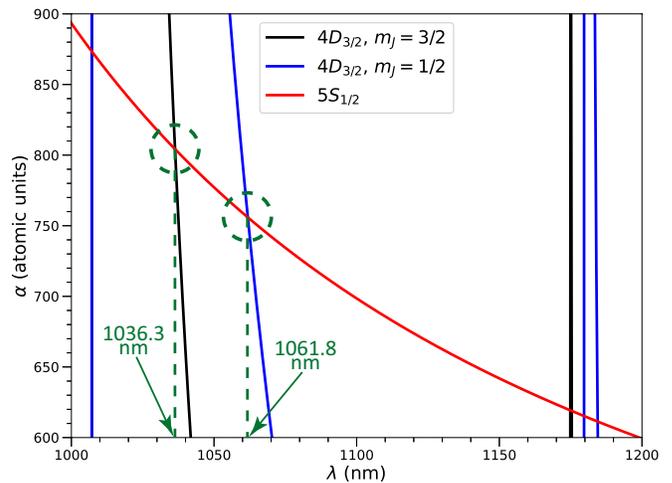}
  \caption{(Color online) Calculated AC polarizability, $\alpha$, as a function of wavelength, $\lambda$, for the Rb 5$S_{1/2}$ and 4$D_{3/2}$ states. Two cases that correspond to ``magic" conditions are indicated with dashed circles. The respective $\lambda$-values are shown as well.} 
  \label{fig5}
\end{figure}

\subsection{Applications of Rb 4$D_J$  states}
\label{subsec:discuss}

We next discuss several applications that can be envisioned for the 4$D_{J}$ state. As mentioned in Sec.~\ref{sec:intro}, the 4$D_J$ states in Rb offer similar features as the 5$D_J$ states, which is attractive for compact atomic-clock designs. Moreover, our calculations~\cite{cardman2021, atoms10040117} of the 4$D_{3/2}$ state's AC polarizability, $\alpha$, shown in Fig.~\ref{fig5} together with $\alpha$ for the ground 5$S_{1/2}$ state, reveal two instances at which $\alpha$ for both states match at NIR wavelengths. One is at 1036.3~nm (for 4$D_{3/2}, m_J = 3/2$), another - at 1061.8~nm (for 4$D_{3/2}, m_J = 1/2$). These cases correspond with ``magic" conditions~\cite{SafronovaRbMagic, Lundblad2010, zhangpra2011, sahoo2013}, at which state-dependent AC shifts match for a given laser wavelength. The HFS is ignored in Fig.~\ref{fig5}, {\sl{i. e.}} the polarizabilities shown are applicable to light-shift traps that are more than about 100~MHz deep. For less-deep traps, the exact magic wavelengths for the  4$D_{3/2}$ HFS states will still be in the same range and can be calculated using methods described in~\cite{Chen2015}. As $\alpha >$ 0 for the ``magic'' cases, it will be possible to trap atoms in both ground and excited states using simple optical-lattice designs during clock-transition interrogation, which helps to diminish motional dephasing and leads to better coherence times~\cite{atomicclocksreview}. Along similar lines, it has been suggested elsewhere to create an off-resonant dipole trap on the 5$S_{1/2} \rightarrow 4D_{5/2}$ one-color two-photon transition~\cite{Roy17} at 1033.3~nm, which is near one of the magic cases in Fig.~\ref{fig5}. Additional state-of-art calculations~\cite{safronova2004, safronova2011} could be used to predict the ``magic" wavelengths with better precision. 
We note that the 5$D_J$ states would be photo-ionized at the aforementioned wavelengths~\cite{duncan2001prameasurement, cardman2021, atoms10040117}.

In a different field of use, the 4$D_J$ states could be employed in Rydberg-atom-based technologies~\cite{Adams_2019} by promoting 4$D$ atoms to Rydberg $P$ or $F$ states with another laser in the range of 698-700~nm. In particular, as the 5$P_J \rightarrow 4D_J$ transitions are driven by telecom-wavelength lasers, further investigations could explore ways to incorporate that into novel Rydberg-atom-based all-optical quantum communication~\cite{Kimble2008}.

\section{Conclusion}
\label{sec:concl}
In summary, we have reported a measurement of the HFS constants of the $^{85}$Rb 4$D_{3/2}$ state using two-photon excitation of cold atoms. The result for the magnetic-dipole constant differs from a previous report~\cite{moon2009} by $>$1\% and lies slightly 
outside the uncertainty overlap between the two measurements. The result for the electric-quadrupole constant agrees with the value given in~\cite{moon2009}. Future high-precision measurements would be of help to address the aforementioned discrepancy. A more sensitive readout method, such as photo-ionization and ion counting, could be utilized for that. Additional studies could be dedicated to measuring the AC polarizability of the 4$D_J$ states at different wavelengths and, in particular, at ones outlined in Fig.~\ref{fig5}. This will aid to assess practical limits for applications of these states in Rydberg-atom physics and novel quantum technologies discussed in this work.

\maketitle
\section*{ACKNOWLEDGMENTS}
We would like to thank the rest of our research group as well as Jordan Lovegrove and Robert Scholten of MOGLabs for useful discussions. This work was supported by the NSF Grant No. PHY-2110049.

\bibliography{references}
\end{document}